# Application of the Theory of Linear Singular Integral Equations to a Riemann Hilbert Problem for a New Expression of Chandrasekhar's H-function in Radiative Transfer.

Rabindra Nath Das*
Visiting Faculty Member , Department of Mathematics , Heritage Institute of Technology , Chowbaga Road , Anandapur, P.O. East Kolkata Township , Kolkata- 700 107 , West Bengal , India.
Email address : rndas1946@yahoo.com

Abstract: The linear non homogeneous singular integral equation( LNSIE) derived from the non-linear non homogeneous integral equation (NNIE) of Chandrasekhar's H –functions is considered here to develop a new form of H-functions .The Plemelj's formulae are applied to that equation to determine a new linear non homogeneous integral equation (LNIE) for H- functions in complex plane. The analytic properties of this new linear integral equation are assessed and compared with the known linear integral equations satisfied by H-functions. The Cauchy integral formulae in complex plane are used to obtain this form of H-functions not dependent on H- function in the integral. This new form of H-function is represented as a simple integral in terms of known functions both for conservative and non-conservative cases. This is identical with the form of H-functions derived by this author by application of Wiener – Hopf technique to LNIE . The equivalence of application of the theory of linear singular integral equation in Riemann Hilbert Problem and of the technique of Wiener- Hopf in linear integral equation in representing the H -functions is therefore established . This new form may be used for solving the problems of radiative transfer in anisotropic and non coherent scattering using the method Laplace transform and Wiener-Hopf technique.

*Permanent address: KB-9, Flat –7, Sector III , Saltlake City (Bidhan nagar) , Kolkata-700098 , West Bengal , India.

## 1 . Introduction :

The H- functions, extensively developed by Chandrasekhar [1] play an important role in the study of radiative transfer in homogeneous plane parallel scattering atmosphere of finite and infinite thickness .. The detailed application of the mathematical theory to the equations of radiative transfer or neutron transport depends , however upon the availability of the H- function in sufficient closed form. This requirement has been partly met by Chandrasekhar [1] . Mathematical framework and the numerical evaluation of these functions have been made by him extensively by his own method. There is still a need for further detailed theoretical and numerical work on these functions , particularly to have a closed form by application of some other methods . Accordingly , we have re -examined the question of mathematical formulation of a new form of the H-functions from a different angle. The object of this is to providing current and future requirements of numerical value of these functions in tabular form for the case of isotropic scattering for any value of particle albedo both in conservative and non conservative cases of physical interest . The work presented in this paper is an extensive formulation of mathematical theory to build up a new form of the H-functions for atmosphere with coherent scattering.

The mathematical form of H functions presented by different authors Chandrasekhar [1] , Kourganoff [2] , Busbridge [3] , Fox [4] , Dasgupta [5],[6] , Das [7] , by different methods are stated to be exact for numerical integration with some suitable constraints. We believe that the forms of H- functions derived so far, as a solution of NNIE or LNIE for numerical computation are in fact dependent on these H- function itself within the integral. However ,we believe that those are worthwhile to frame a basis for a new functional representation of H – functions as a solution of the LNIE of H- functions .

The forms of H-functions derived by Fox[4], Zelazny [8] , Mullikin[9] , Zweifel[10],Ivanov[11] ,Siewert [12], Das [13] are considered to be important from theoretical and numerical point of view . Those solutions are not dependent on the H-function itself.

The forms of H – functions derived by Siewert [14] , Garcia and Siewert [15],Sulties and Hill [16] , Barichello and Siewert [17 , 18] , Bergeat and Rutily [ 19,20] in terms of new functions from different standpoints are still considered to be useful for numerical and theoretical point of view.

Further more we believe that the result obtained by this method of application of the theory of singular integral equation in this work to a LNSIE of H functions for solving a Riemann Hilbert problem is a new and sufficiently different from others in the literature of H- functions to warrant its communication. Consequently we hope that this functional representation of H – functions extends the result of Fox [4] for better understanding of the method used by him. We hope that these results identify the equivalence of the application of Wiener-Hopf technique to LNIE of H-functions Das [13] and the application of theory of linear singular integral equation to LNSIE of H- functions so as to have the same form of H – function by two different approaches. The existence and uniqueness of those functions are achieved already in Das[13]. It is therefore shown that it is now safe to handle the LNSIE or LNIE of H- functions with due respect to the concluding comment of Busbridge [21].

## 2. Mathematical analysis :

The integro-differential equations of radiative transfer have been solved by different authors to obtain the intensity of radiation at any optical depth. and at the boundary. They finally derived solutions in terms of Chandrasekhar [1] H-functions.

The H –functions satisfy NNIE as

$$H(x) = 1 + x\, H(x) \int_0^1 U(u)\, H(u)\, du / (u + x), \quad 0 \le x \le 1, \quad (1)$$

where U(u) is a known characteristic function. In physical context to solve equation (1), certain restrictions on U(u) are necessary :

$$\text{i)}\ U_0 = \int_0^1 U(u)\, du = \tfrac{1}{2} \quad (2)$$

where $U_0 = \tfrac{1}{2}$ refers to conservative case and $U_0 < \tfrac{1}{2}$ refers to non conservative case ;

ii) U(u) is continuous in the interval (0,1) and satisfy Holder conditions, viz.,

$$I\ U(u_2) - U(u_1)\ I\ <\ K\ \ I\ u_2 - u_1\ I^\alpha \quad (3)$$

where $u_2, u_1$ lie in the interval (0,1) ($u_1$ is in the neighbourhood of $u_2$), where K and α are constants and $0 < \alpha \le 1$.

If $z = x + iy$, a complex variable in the complex plane cut along (-1,1) then Chandrasekhar [1] proved that

$$(H(z)\, H(-z))^{-1} = 1 - 2z^2 \int_0^1 U(u)\, du / (z^2 - u^2) = T(z), \quad (4)$$

and he determined H – function to satisfy the integral solution as

$$\log H(z) = (2\pi i)^{-1} \int_{-i\infty}^{i\infty} z \log(T(w)) \, dw / (w^2 - z^2) \tag{5}$$

where $T(z)$ has the properties:

i) $T(z)$ is analytic in the complex z plane cut along $(-1,1)$; ii) it is an even function of $z$; iii) it has two logarithmic branch points at $-1$ and $+1$; iv) $T(z) \to 1$ as $|z| \to 0$, v) it has only two simple zeros at infinity when $U_0 = \frac{1}{2}$; vi) only two real simple zeros at $z = +1/k$ and $z = -1/k$ where k is real, $0 < k \leq 1$ when $U_0 < \frac{1}{2}$; vii) $T(z) \to -C/z^2$ as $z \to \infty$ when $U_0 = \frac{1}{2}$, $C = 2 U_2$, a real positive constant; viii) $T(z) \to D = 1 - 2 U_0$ as $z \to \infty$ when $U_0 < 1/2$, D is a real positive constant; ix) it is bounded on the entire imaginary x) $T(\infty) = D = 1 - 2 U_0$ when $U_0 < 1/2$

$$= (1 - \int_0^1 H(x) U(x) \, dx)^2 = 1/(H(\infty))^2$$

where $U_r = \int_0^1 u^r U(u) \, du, \quad r = 1,2,3\ldots$ \tag{7}

Fox (1961) determined that, equation (1) is equivalent to LNIE in the complex z plane cut along $(-1.1)$

$$T(z) H(z) = 1 + z \int_0^1 U(u) H(u) \, du / (u - z) = 1/H(-z) \tag{8}$$

with constraints

i) when $U_0 < 1/2$

$$1 + \int_0^1 U(u) H(u) \, du / (u k - 1) = 1/H(-1/k) = 0 \tag{9}$$

$$1 + \int_0^1 U(u) H(u) \, du / (u k + 1) = 1/H(1/k) = 0 \tag{10}$$

and ii) when $U_0 = 1/2$

$$1 - \int_0^1 U(u) H(u) \, du = 0 \tag{11}$$

and to LNSIE if real$(z) = x$ lies in $(-1,1)$

$$T_0(x) H(x) = 1 + P x \int_0^1 U(u) H(u) \, du / (u - x) \tag{12}$$

where P represents the Cauchy Principal value of the integral and .

$$T_0(t) = 1 - t \int_0^1 U(x)\,dx/(x+t) +$$

$$t \int_0^1 (U(x) - U(t))\,dx/(x-t) + t\,U(t)\ln((1-t)/t). \tag{13}$$

In this paper, we shall consider the LNSI Eq (12) of H-functions to solve it by the theory of linear singular integral equation fully described in the standard work of Muskhelishvilli [22] associated with the theory of analytic continuation and Cauchy theory on analytic function in complex plane . We derive a unique solution which is obtained by Das[23] by using Wiener Hopf technique along with theory of analytic continuation and Cauchy theory of analytic function in complex plane . We shall show the equivalence of Riemann Hilbert Theory for solving LNSIE and the Wiener Hopf technique for solving LNIE so far as it relates to determination of a new form for H- functions

Let us consider the function

$$\Phi(z) = \int_L \psi(t)\,dt/(t-z), \tag{14}$$

where L is some contour in the complex z plane with a cut along (0,1) and z does not lie on L . Here $\Phi(z)$ be an analytic function of z in the complex plane cut along (0,1) and be $O(1/z)$ when $z \to \infty$. If z is on L , then the integral for $\Phi(z)$ is taken to be a Cauchy principal value and the curve L becomes a line of discontinuity for the function $\Phi(z)$., The function $\Phi(z)$ must pass through a discontinuity when z crosses L. Let A and B be the end points of the arc L at z = 0 and z = 1 respectively .

We want to determine the value of $\Phi(z)$ as $z \to t_0$ from both sides of the contour where $t_0$ lie on L and $t_0$ is not an end point of L. We define the region $D^+$ as the region to the left of the contour( if we look to the left moving counter clockwise along the contour) and $D^-$ as the region to the right .

.Let us define

$$\Phi^{+ \text{ or } -}(t_0) = \lim_{z \to t_0} \Phi(z), \quad z \text{ belongs to } D^{+ \text{ or } -} \qquad (15)$$

We also define Cauchy principal value integral

$$\Phi(t_0) = P \int_L \psi(t) \, dt / (t - t_0) \qquad (16)$$

$$= \lim_{\epsilon \to 0} \int_{L-L_\epsilon} \psi(t) \, dt / (t - t_0) \qquad (17)$$

where $L_\epsilon$ is that part of L contained within a circle of radius $\epsilon$ with center at $t = t_0$ and where $L - L_\epsilon$ is that part of L contained without it.

We assume $\psi(t)$ is analytic at $t = t_0$ and continuous every where, hence by analytic continuation $\psi(t)$ is analytic in a small neighbourhood of $t_0$ and this continuation can be extended to the whole complex plane.

We therefore using the definition in equations (16) and (17)

$$\Phi^+(t_0) = \lim_{\substack{z \to t_0 \\ \epsilon \to 0}} \Phi(z), \quad z \text{ lie within } D^+$$

$$= \Phi(t_0) + i\pi \psi(t_0) \qquad (18)$$

$$\Phi^-(t_0) = \lim_{\substack{z \to t_0 \\ \epsilon \to 0}} \Phi(z), \quad z \text{ lie within } D^-$$

$$\Phi^-(t_0) = \Phi(t_0) - i\pi \psi(t_0) \qquad (19)$$

Equation (18) and (19) are referred to as the Plemelj formulae. They may be equivalently written as

$$\varphi^+(t_0) - \Phi^-(t_0) = 2\pi i \psi(t_0) \qquad (20)$$

$$\Phi^+(t_0) + \Phi^-(t_0) = 2 P \int \psi(t) \, dt / (t - t_0) \qquad (21)$$

The end points of the contour L cause considerable difficulty. A complicated analysis in Muskhelishvilli (1953) shows that at the end points $t_A$ and $t_B$ of L, $\Phi(t)$ will behave as

$$|\Phi(t)| \leq K_A / |t - t_A|^{\alpha_A}, \quad 0 \leq \alpha_A < 1 \qquad (22)$$

$$|\Phi(t)| \leq K_B / |t - t_B|^{\alpha_B}, \quad 0 \leq \alpha_B < 1 \qquad (23)$$

where $K_A$, $K_B$, $\alpha_A$ and $\alpha_B$ are constants.

### 3. Determination of a new LNIE for H-functions :

we shall now introduce a function $Y(z)$ of the complex variable z

$$Y(z) = \int_0^1 U(t) H(t) \, dt / (t - z) \qquad (24)$$

which has the properties :

i) $Y(z)$ is analytic in the complex plane cut along (0,1) ; ii) $Y(z) = O(1/z)$ when $z \to \infty$ ; iii) $|Y(z)| \leq K_0 / |z|^{\alpha_0}$, $0 \leq \alpha_0 < 1$ ; iv) $|Y(z)| \leq K_1 / |1-z|^{\alpha_1}$, $0 \leq \alpha_1 < 1$; v) $U(t) H(t)$ is continuous in the interval (0,1) and satisfy equivalent Holder conditions stated in equation (3) and $K_0, K_1, \alpha_0$ and $\alpha_1$ are constants.

We shall now apply the Plemelj's formulae (20) and (21) to the equation (12) to have

$$Y^+(x) = g(x)Y^-(x) + 2\pi i \, U(x) / T^-(x), \quad 0 \leq x \leq 1 \qquad (25)$$

where $Y^+(x)$, $Y^-(x)$ and $g(x)$ are complex numbers and $g(x)$ is related to

$$T^+(x) = T_0(x) + i\pi x U(x) \qquad (26)$$

$$T^-(x) = T_0(x) - i\pi x U(x) \qquad (27)$$

$$g(x) = (T_0(x) + i\pi x U(x)) / (T_0(x) - i\pi x U(x))$$

$$= T^+(x)/T^-(x) \qquad (28)$$

where $T_0(x)$ is given by equation (13).

From equation (25) we have to determine the unknown function $Y(z)$ for being analytic in the complex plane cut along (0,1) having properties outlined above and $g(x)$ and $U(x)$ are known functions. This is known as non homogeneous Riemann Hilbert problem.

We shall first determine the solution of homogeneous Riemann Hilbert problem obtained from equation (25).

It is that of finding a new analytic function X(z) for which

$$X^+(x) = g(x)\, X^-(x), \quad 0 \leq x \leq 1 \tag{29}$$

where X(z) is to satisfy the following properties:

i) X(z) is analytic in the complex plane cut along (0,1) ; ii) X(z) is not zero for all z in complex plane cut along (0,1) ; iii) O(1/z) when z → ∞ ; iv) $|X(z)| \leq K_2 / |z|^{\alpha_2}$, $0 \leq \alpha_2 < 1$ ; and v) $|X(z)| \leq K_3 / |1-z|^{\alpha_3}$, $0 \leq \alpha_3 < 1$, where $K_2, K_3, \alpha_2$ and $\alpha_3$ are constants.

We now assume that for $0 \leq x \leq 1$, U(x) is real, positive, single valued and satisfy Holder condition in (0,1). We also assume that $T_0(x)$ is not equal to zero both in the conservative case $U_0 = \frac{1}{2}$ and in non conservative cases $U_0 < \frac{1}{2}$. As $T_0(x)$ is dependent on U(x) it can be proved that $T_0(x)$ is real, one valued and satisfy Holder conditions in (0,1) and modulus of ($T_0(x) + i\pi x U(x)$), ($T_0(x) - i\pi x U(x)$) are not equal to zero, $T_0(0) = 1$ and $T_0(1) \to -\infty$ as z → 1 from within the interval (0,1).

Taking logarithm to equation (29) and using (28) we find that

$$\log(X^+(x)) - \log(X^-(x)) = \log g(x) = 2 i \theta(x) \tag{30}$$

where

$$\tan \theta(x) = \pi x\, U(x) / T_0(x) \; ; -\pi/2 < \theta(x) < \pi/2 \tag{31}$$

$$\theta(0) = 0, \tag{32}$$

and θ(x) is assumed single valued.

We shall evaluate X(z). Using Plemelj's formulae (20) to equation (30) we get

$$\log X(z) = (2\pi i)^{-1} \int_0^1 \log g(u)\, du /(u - z) + P_n(z) \tag{33}$$

where $P_n(z)$ is an arbitrary polynomial in z of degree n. in the complex plane cut along (0,1) and it is continuous in the complex plane across a cut along (0,1) because it is analytic there. Thus equation (33) does also satisfy the first of Plemelj's formulae (20).

We set

$$L(z) = (2\pi i)^{-1} \int_0^1 \log g(u)\, du / (u - z) \tag{34}$$

Equation (33) with equation (34) then yields

$$X(z) = X_0(z) \exp(L(z)) \qquad (35)$$

where $X_0(z) = \exp(P_n(z))$ is analytic for all $z$ in the complex plane and is to be determined such that $X(z)$ given by equation (35) satisfy all properties outlined above and the use of the Plemelj's formulae is not invalidated.

We have to determine the value of $X_0(z)$ at the end points: Equation (34) may be written as

$$L(z) = (2\pi i)^{-1} \log g(z) \int_0^1 du/(u-z)$$

$$+ (2\pi i)^{-1} \int_0^1 (\log g(u) - \log g(z)) \, du/(u-z) \qquad (36)$$

At the end point $z = 0$, equation (36) can be written as

$$L(z) = = -(\log g(z)/2\pi i) \log |z| + q_0(z) \qquad (37)$$

where $q_0(z)$ incorporates the second integral of equation (36) and the value of first term at $z = 1$. Here $q_0(z)$ is clearly a bounded function. We therefore find that, in the neighborhood of the end point at $z = 0$, the behavior of $X(z)$ is dominated by

$$X(z) \to X_0(z) \, |z|^{-(\log g(z)/2\pi i)} \qquad (38)$$

Near the end point $z=0$ for the Plemelj' formulae to be valid, $X_0(z)$ and $X(z)$ must behave in the neighborhood of $z = 0$ as

$$X_0(z) \to |z|^{m_0} \qquad (39)$$

$$X(z) \to |z|^{(m_0 - (\log g(z)/2\pi i))} \qquad (40)$$

We must have the constant $m_0$ to satisfy there

$$0 \le \text{Real}(\log g(z)/2\pi i) - m_0 < 1 \qquad (41)$$

As $X_0(z)$ will be analytic and continuous across the cut

in complex $z$ plane along $(0,1)$ we must have using eqs.(30),(32)

$$m_0 = \text{Real}(\log g(0)/2\pi i) = \theta(0)/\pi = 0 \qquad (42)$$

At the end point $z = 1$, equation (36) can be written as

$$L(z) = = (\log g(z)/2\pi i) \log |1-z| + q_1(z) \qquad (43)$$

where $q_1(z)$ incorporates the second integral of equation (36) and the value of first term at $z = 0$. Here $q_1(z)$ is clearly a bounded function. Hence, in the neighborhood of the end point at $z = 1$, the behavior of $X(z)$ will be dominated by $X_0(z)$ as

$$X(z) \rightarrow X_0(z) \; |1-z|^{(\log g(z) / 2\pi i)} \qquad (44)$$

and for the Plemelj' formulae to be valid there, $X_0(z)$ and $X(z)$ must also behave in the neighborhood of $z = 1$ as

$$X_0(z) \rightarrow |1-z|^{m_1} \qquad (45)$$

$$X(z) \rightarrow |1-z|^{(m_1 + (\log g(z) / 2\pi i))} \qquad (46)$$

We must have the constant $m_1$ to satisfy there

$$0 \leq - \text{Real}(\log g(z) / 2\pi i) - m_1 < 1 \qquad (47)$$

As $X_0(z)$ will be analytic and continuous across the cut in complex z plane along $(0,1)$ hence using eq.(30)

$$m_1 = - \text{Real}(\log g(1) / 2\pi i) = -\theta(1) / \pi \qquad (48)$$

But we know that

$$\theta(1) = r\pi, \quad r = 1,2,3 \text{ etc} \qquad (49)$$

so $m_1 = -1$ for $r = 1 = \frac{1}{2}$ the zeros of $T(z)$ in the complex plane cut along $(-1,1)$. Hence on consideration of the properties of $X_0(z)$ at both end points we have

$$X_0(z) = (1-z)^{-1} \qquad (50).$$

Hence equation (35) using equations(50), yields

$$X(z) = (1-z)^{-1} \exp(L(z)) \qquad (51).$$

This is the solution of homogeneous Hilbert problem mentioned in equation (29) in the complex z plane cut along $(0,1)$.

This can be written in explicit form when z is in complex in z plane cut along $(0,1)$ using eqs.(34) & (30)

$$X(z) = (1-z)^{-1} \exp\left( (2\pi i)^{-1} \int_0^1 \log g(u) \, du / (u - z) \right) \qquad (52)$$

$$= (1-z)^{-1} \exp\left(\pi^{-1} \int_0^1 \theta(u)\, du/(u-z)\right) \quad (53)$$

where $\theta(u)$ is given by using equation (31) as

$$\theta(u) = \tan^{-1}(\pi u\, U(u)/T_0(u)) \quad (54)$$

We shall determine a solution of non homogeneous Hilbert problem. The equation (25) may be changed by first writing

$$X^+(u) = g(u)\, X^-(u), \quad 0<u<1 \quad (55)$$

when u lie on cut along (0,1), $X^+(u), g(u)$ and $X^-(u)$ are all nonzero. Substitution of equation (55) in equation (25) we get

$$Y^+(u)/X^+(u) - Y^-(u)/X^-(u) = 2\pi i\, U(u)/X^+(u)\, T^-(u), \quad (56)$$

when u lie in the interval (0,1). Using Plemelj's formulae in equation(20) to equation (56) we get for z in complex plane cut along (0,1),

$$Y(z) = X(z) \int_0^1 U(u)\, du/(X^+(u)\, T^-(u)(u-z))$$

$$+ P_m(z)\, X(z) \quad (57)$$

where $P_m(z)$ is an arbitrary polynomial in z of degree m and is continuous in the complex plane across a cut along (0,1).

Thus equation (57) does satisfy the first of Plemelj's formulae as in. equation (20) The arbitrariness of $P_m(z)$ is removed usually by examining the behaviour of Y(z) and X(z) at infinity and /or the end points of the cut along (0,1). Knowledge of X(z) then completes the solution for Y(z). Y(z) will then be the sum of solution of homogeneous Hilbert problem in equation (29) and the particular solution of the equation (56).

We now return our attention to the evaluation of the particular solution of equation (56)for Y(z) from equation (57) and in particular, to the determination of the polynomial $P_m(z)$. Since X(z) is equivalent to $\exp(L(z))$, the term appearing on the RHS of equation (57) is a polynomial of degree m. However, if we use the properties of X(z) and Y(z) when z→ ∞, we must have

$$P_m(z) = 0 \quad (58)$$

Hence we get the particular solution $Y_p(z)$ of equation (56)in complex z plane cut along (0,1)) for Y(z) as

$$Y_p(z) = X(z) \int_0^1 U(u) \, du / (X^+(u) T^-(z) (u - z))  \quad (59).$$

We shall now represent $Y_p(z)$ in terms of equation $X(z)$ in the complex z plane cut along (0,1) by using Cauchy's integral theorems

We set

$$V(z) = (2\pi i)^{-1} \int_0^1 U(u) du / (X^+(u) T^-(u) (u-z))  \quad (60)$$

where $V(z)$ is analytic in the complex z plane cut along (0, 1).

We consider two contours as follows:

$$F(z) = (2\pi i)^{-1} \int_{C_1} F(u) \, du / (u-z)  \quad (61)$$

where $C_1 = L_1 \cup L_2$. Here contour $L_1$ is a sufficiently large in the complex z plane to contain the cut along (0,1) within and $L_2$ is a circle with center at the origin and of very large radius R and $L_1$ lies interior to this $L_2$ and z lies outside $L_1$ but inside $L_2$. Both $L_1$ and $L_2$ are taken in counter clock wise sense. The function $X(z)$ and $1/X(z)$ are analytic and nonzero in the annulus $C_1$. We shall now apply the Cauchy Integral theorem on the function $F(z) = 1/zX(z)$ around $C_1$ to obtain

$$1/zX(z) = (2\pi i)^{-1} \int_{L_2} dw / (X(w) w (w-z))$$

$$- (2\pi i)^{-1} \int_{L_1} dw / (X(w) w (w-z))  \quad (62)$$

Using equation (51) we can write

$$1/X(w) = (1 - w) \exp(-L(w))  \quad (63)$$

and when $w = u$, u real and $0 < u < 1$

we can write from eq.(63)

$$1/X^+(u) = (1 - u) \exp(-L^+(u))  \quad (64)$$

$$1/X^-(u) = (1 - u) \exp(-L^-(u))  \quad (65)$$

where the superscript '+' denotes the value from above the cut along (0,1) and the superscript '-' denotes the value from below the cut along (0,1) of the respective functions in the complex z plane and

$$L^+(u) = (\pi^{-1} P \int_0^1 \theta(t) \, dt / (t-u) + i\theta(u)) \qquad (66)$$

$$L^-(u) = (\pi^{-1} P \int_0^1 \theta(t) \, dt / (t-u) - i\theta(u)) \qquad (67)$$

Using eqs. (66) & (67) in equations (64) and (65) we can write that

$$1/X^+(u) - 1/X^-(u) = (1 - \exp(2i\theta(u))) / X^+(u) \qquad (68).$$

Using equation (29) and (30) we get

$$\exp(2i\theta(u)) = T^+(u) / T^-(u) \qquad (69).$$

Using equations (69), (26) & (27) in equation (68) we get

$$1/X^+(u) - 1/X^-(u) = -2\pi i \, u \, U(u) / T^-(u) X^+(u) \qquad (70)$$

To make the contour $L_1$ to be well defined we shall shrink the contour $L_1$ to i) a circle $C_0$ around the origin of the complex plane in counter clockwise sense such that $w = r \exp(i\alpha)$, $0 \le \alpha \le 2\pi$; ii) a line CD below the cut along (0,1) from r to 1-r where $w = u$, real, u varies from 0 to 1; iii) a circle $C_2$ counter clockwise sense around $w=1$ such that $w = 1 + r \exp(i\beta)$, $-\pi \le \beta \le \pi$ and iv) a line BA above the cut along (0,1) from 1-r to r on which $w = u$, real, u varies from 1 to 0.. The value of the integral, on the circle $C_0$, in the limit $r \to 0$ becomes

$$(2\pi i)^{-1} \int_{C_0} dw / (X(w) w (w-z)) = -(z X(0))^{-1} \qquad (71)$$

The value of the integral, on the circle $C_2$, in the limit $r \to 0$ becomes

$$(2\pi i)^{-1} \int_{C_2} dw / (X(w) w (w-z)) = 0 \qquad (72)$$

The value of the integral, on the line CD, in the limit $r \to 0$ becomes

$$(2\pi i)^{-1} \int_{CD} dw / (X(w) w (w-z)) = (2\pi i)^{-1} \int_0^1 du / (X^-(u) u (u-z)) \qquad (73)$$

The value of the integral, on the line BA, in the limit $r \to 0$ becomes

$$(2\pi i)^{-1} \int_{BA} dw / (X(w) w (w-z)) = (2\pi i)^{-1} \int_1^0 du / (X^+(u) u (u-z)) \quad (74)$$

Hence the integral, on the contour $L_1$ in eq(62), becomes using eqs.(70-74)

$$(2\pi i)^{-1} \int_{L_4} dw /(X(w) w (w-z)) = -(2\pi i)^{-1} \int_0^1 (1/X^+(u) - 1/X^-(u)) du /(u (u-z))$$
$$- 1 / (z X(0)) \qquad (75)$$

$$= \int_0^1 U(u) du /(X^+(u) T^-(u) (u-z))$$

$$- 1 / (z X(0)) \qquad (76)$$

The integral, on the contour $L_2$, when $R \to \infty$ becomes

$$(2\pi i)^{-1} \int_{L_2} dw /(X(w) w (w-z)) = -1 \qquad (77)$$

as $z X(z) = -1$ when $z \to \infty$ \qquad (78)

Hence equation (60) with equations (62), (76) and (77) gives

$$V(z) = (z X(0))^{-1} - 1 - (z X(z))^{-1} \qquad (79)$$

Hence the particular solution $Y_p(z)$ of equation (56) gives

$$Y_p(z) = X(z) V(z) = X(z) / (z X(0)) - X(z) - 1/z \qquad (80)$$

Hence the general solution of the equation (25) in the complex z plane, is

$$Y(z) = A X(z) + Y_p(z) \qquad (81)$$

where A is a constant yet to be determined by using the pole of Y(z), if any, at the zero of T(z) and X(z), $Y_p(z)$ are determined by equations(51) and (80) respectively. This Y(z) in Equation (81) will help in representing 1/H(-z) in equation (8).

We shall now show that H(z) in equation (8) is continuous across the cut along (0,1).. Let H(x) be any real valued solution of NNIE at equation (1) for $0 \le x \le 1$ then it can be proved that H(z) defines in equation (8) is the meromorphic extension of H(x), $0 \le x \le 1$ to the complex domain 1z1>0. In addition H(x) will satisfy the LNSIE given by equation (12). By using the Plemelj's first formulae as in equations(18) &(19) to equation (8) when z approaching the cut along (0,1) from above and below respectively :

$$T^+(x) \, H^+(x) = 1 + x \, P\int_0^1 H(t) \, U(t) \, dt / (t-x) + i\pi \, H(x) \, U(x), \quad 0 \le x \le 1 \quad (82)$$

$$T^-(x) \, H^-(x) = 1 + x \, P\int_0^1 H(t) \, U(t) \, dt / (t-x) - i\pi \, H(x) \, U(x), \quad 0 \le x \le 1 \quad (83)$$

where superscript '+' and '-' denotes the value as z approaches from above and below to the cut along (0,1). Using the values of $T^+(x)$ and $T^-(x)$ from equations (26), (27) to equations (82) and (83) and using the equation (12) therein we get

$$H^+(x) = H(x) = H^-(x) \quad (84)$$

This shows that H(z) defined by equation (8) is continuous across the cut along (0,1) and real valued on (0,1) Hence it indeed defines a meromorphic extension of H(x) at least in the region $0 < \text{Re}(z) < 1$, $\text{Im}(z) > 0$ and $\text{Im}(z) < 0$ and $T(1) \ne 0$, and it can be said that z =1 is a removable singularity. We shall now frame a new LNIE from the theory of linear singular integral equations. Using equations (8),(26),(27),(81),(82),(83) & (84) in equation (24), we get,

$$(T^+(x) \, H(x) - x \, A \, X^+(x) - x \, Y_p^+(x))$$
$$- (T^-(x) \, H(x - x \, X^-(x) - x \, Y_p^-(x))) = 0 \quad (85)$$

Using first of Plemelj's formulae as in equation (20) to equation (85) we get

$$T(z) \, H(z) - z \, A \, X(z) - z Y_p(z) = P_p(z) \quad (86)$$

where $P_p(z)$ an arbitrary polynomial is to be determined from the properties of X(z), Yp(z), T(z) and H(z) at the origin..

Since H(z) and T(z) both → 1 when z→0, equation (86) gives

$$P_p(z) = 1 \quad (87)$$

Equation (86) with equation (87) gives **a new representation** of LNIE in the complex z plane cut along (0,1) from the theory of linear singular integral equation and by analytic continuation to complex plane cut along (-1,1) to

$$T(z) \, H(z) = 1 + z \, A \, X(z) + z \, Y_p(z) \quad (88).$$

## 4. Determination of constant :

Equation (88) with equation (8) and (81) gives

$$1/H(-z) = 1 + z A X(z) + z Y_p(z) \tag{89}$$

$$= 1 + z Y(z) \tag{90}$$

We have now to determine the constant A in equation (88). We know that $T(z)$ has two zeros at $z = 1/k$ or $-1/k$, $0 \leq k \leq 1$. $1/H(-z)$ has a zero at $z = 1/k$. Equation (89), on substitution of $z = 1/k$, gives

$$1 + A X(1/k) / k + Y_p(1/k) / k = 0 \tag{91}$$

Using equation (80) for $z = 1/k$ we get

$$Y_p(1/k) = k ( X(1/k)/X(0) - X(1/k)/k - 1) \tag{92}$$

Equation (91) and (92) we get

$$1 + A X(1/k)/k + X(1/k)/X(0) - X(1/k)/k - 1 = 0 \tag{93}$$

This equation (93) on simplification gives

$$A = 1 - k / X(0) \tag{94}$$

We shall have to determine $X(0)$. Substituting $z=0$ in equation (53) we get

$$X(0) = \exp( \pi^{-1} \int_0^1 \theta(t) d t/t ) \tag{95}$$

We can determine other form of A for a new form of $X(0)$. In equation (24),(51) & (59) if we make $z \to \infty$, then

$$Y(z) = - \beta_0 / z^1 - \beta_1 / z^2 - \beta_2 / z^3 -- \tag{96}$$

$$X(z) = - 1/z - O(1/z^2) ; \tag{97}$$

$$Y_p(z) = O(1/z^2) ; \tag{98}$$

$$\text{where } \beta_r = \int_0^1 x^r U(x) H(x) d x , r = 0,1,2,3.. \tag{99}$$

Equating the like power of $z^{-1}$ from both sides

of equation (81) for large z we get after manipulation with eq(6)

$$A = \beta_0 = \int_0^1 U(x) H(x) d x = 1 - (D)^{1/2} \tag{100}$$

From equations (94),(100), we get

$$X(0) = k / D^{1/2} \qquad (101)$$

From eqs.(101), (95) we can determine

$$\theta_{-1} = \pi^{-1} \int_0^1 \theta(t) dt/t = \ln( k / D^{1/2} ) \qquad (102)$$

## 5. Determination of new form of H- function :

We shall now determine the new form of H(z). On substitution of the values of A, X(0) and $Y_p(z)$ from eqs.(94),(101) and (80) respectively in equation (89) we get in the complex plane cut along (0,1)

$$1/ H(-z) = (1 - kz) X(z) (D)^{1/2} / k , \qquad (103)$$

where X(z) is given by equation (53)

Hence in the complex z plane cut along (-1,0) we get

$$H(z) = k /( X(-z)( 1 + k z ) ( D)^{1/2} ). \qquad (104)$$

Using the values of X(z) and X(-z) in equations (103) and (104)

we get the new form of H(z) and H(-z) as

$$H(z) = k ( D)^{-1/2} (1+z) \exp( - \pi^{-1} \int_0^1 \theta(u) du /( u + z) ) /(1 + k z ). \qquad (105)$$

$$H(-z) = k ( D)^{-1/2} (1-z) \exp( - \pi^{-1} \int_0^1 \theta(u) du /( u - z) ) /(1 - k z ). \qquad (106)$$

These are the same form derived by Das [13] using Wiener Hopf technique to linear non homogeneous integral equation in Eq.(8).

## 6. New form of H- function in Cauchy Principal value sense:

We have to find a new form for H(x) in Cauchy's principal value sense by application of Plemelj's formula. We know from equation ( 81 ) using eq.(94) that

$$Y(z) = X(z) (1-kz) / ( z X(0) ) \qquad (107)$$

Using first of the Plemelj's formulae (20) to equation (107) we get

$$H(x) = ( Y^+(x) - Y^-(x) ) / ( 2\pi i \, U(x) ) \qquad (108)$$

$$= (1 - kx)(X^+(x) - X^-(x)) / (2\pi i \, x \, U(x) \, X(0)) . \quad (109)$$

But

$$X^+(x) - X^-(x) = 2i \sin(\theta(x)) \exp(P \, \pi^{-1} \int_0^1 \theta(t) \, dt / (t - x)) / (1-x) , \quad (110)$$

$$\sin(\theta(x)) = \pi \, x \, U(x) / (T_0(x) + \pi^2 x^2 U^2(x))^{1/2} \quad (111)$$

$$X(0) = k \, D^{-1/2} \quad (112)$$

$$D = (1 - 2 U_o) \quad (113)$$

Equation (109) with equations (110-113) gives

$$H(x) = (1 - kx) \, D^{1/2} \, \exp(-\pi^{-1} P \int_0^1 \theta(t) \, dt / (t - x)) \, X$$

$$(T_0(x) + \pi^2 x^2 U^2(x))^{-1/2} (k(1-x))^{-1} \quad (114)$$

This is the new explicit form of H(x) different from Fox's [4]solution . The integral is Cauchy principal Value sense.

### 7. Removal of Cauchy Principal Value sense from H- function:

We shall now show that the representation of H(x) in eq.(114) is nothing but the representation of H(x) in equation .(105). It can be done by way of removal of singular part from equation (114).

$$\text{Let} \quad M(x) = \pi^{-1} P \int_0^1 \theta(t) \, dt / (t - x) \quad , 0 \le x \le 1 \quad (115)$$

$$N(x) = \pi^{-1} \int_0^1 \theta(t) \, dt / (t + x) \quad , 0 \le x \le 1 \quad (116)$$

If z approaches from above to the cut along (0,1) to x , $0 \le x \le 1$ we get

$$H^+(x) \, H^+(-x) = 1 / T^+(x) \quad (117)$$

Here $H^+(x) = H(x)$ as H(x) is continuous across the cut along (0,1) but $H^+(-x)$ is not continuous there . Therefore we get $H^+(-x)$ , $H^+(x)$ from equation (106) & (105) using equations (115) & (116) that

$$H^+(-x) = (1-x) \, k \, (D)^{-1/2} \exp(-M(x) - i\pi \, \theta(x)) / (1-kx) \quad (118)$$

$$H^+(x) = (1-x) \, k \, (D)^{-1/2} \exp(-N(x)) / (1+kx) \quad (119)$$

$$T^+(x) = T_0(x) + i \pi x U(x) \qquad (120)$$

Using equations (118-120) in equation (117) we get

$$(1 - x^2) k^2 D^{-1} (1 - k^2 x^2)^{-1} \exp(-M(x) - N(x)) (\cos \theta(x) - i \sin \theta(x))$$

$$= (T_0(x) - i\pi x U(x)) / (T_0(x) + \pi^2 x^2 U^2(x)) \qquad (121)$$

Equating the real and imaginary parts of equation (141) we get

$$(1 - x^2) k^2 D^{-1} (1 - k^2 x^2)^{-1} \exp(-M(x) - N(x)) \cos \theta(x)$$

$$= T_0(x) / (T_0(x) + \pi^2 x^2 U^2(x)) \qquad (122)$$

$$(1 - x^2) k^2 D^{-1} (1 - k^2 x^2)^{-1} \exp(-M(x) - N(x)) \sin \theta(x)$$

$$= \pi x U(x) \} / (T_0(x) + \pi^2 x^2 U^2(x)) \qquad . \qquad (123)$$

But using equations (111), (115) and (116) in equation (123) we get

$$(1 - x^2) k^2 D^{-1} (1 - k^2 x^2)^{-1} \exp(-M(x) - N(x))$$

$$= 1 / (T_0(x) + \pi^2 x^2 U^2(x))^{1/2} \qquad . \qquad (124)$$

Now we use equation (124) to equation (114) in order to eliminate the Principal part of the integral in equation (114) and we get the same form of H(x) ( as in equation (105) ) with constraints (9) & (10) in non conservative cases and constraint (11) in conservative cases

Using equations (101) and (95) to equation (105) we get the form outlined by Mullikin [9].in non conservative cases:

$$H(z) = (1+z) \exp\left( \pi^{-1} \int_0^1 \theta(u) \, du / u(u + z) \right) / (1 + k z). \qquad (125)$$

### 8. Determination of H- function in conservative cases :

From equation (105) we can derive the representation of H(z) for conservative cases. In conservative cases $U_0 = 1/2$, T(z) will have zeros at infinity. Hence $k = 0$. When k=0, $U_0 = 1/2$, D = 0 by equation (113). Hence when k→0, k $D^{-1/2}$ → $(2 U_2)^{-1/2}$, in conservative case, H(z) in equation (105) will take the form

$$H(z) = (2 U_2)^{-1/2} (1+z) \exp\left( -\pi^{-1} \int_0^1 \theta(u) \, du / (u + z) \right) . \qquad (126)$$

with constraint in equation (11).

## 9. Admissible solutions:

Chandrasekhar [1], Busbridge [3], Dasgupta [23], Abhyankar [24] proved that, in non conservative cases, these two constraints(9) &(10) will provide two different unique solutions but in conservative cases those two solutions will be identical to form one unique solution with constraint(11). In non conservative cases, they also derived the form of constraints from equations (9) and (10) respectively to

$$\int_0^1 U(u) H(u) \, du = 1 - D^{1/2} \qquad (127)$$

$$\int_0^1 U(u) H(u) \, du = 1 + D^{1/2} \qquad (128)$$

They took one meaningful solution of H(z) of equation (8) for physical context with constraint (127). The other solution $H_1(z)$ with constraints (128) was

$$H_1(z) = (1+kz) \, H(z) /(1- kz) \qquad (129)$$

We therefore, resolve that, in non conservative cases, equation (105) for H(z) with constraints (9) or (127) will provide meaningful unique solution of equation (12) of physical interest and the other unique solution $H_1(z)$ will be defined by equation (129) with H(z) from equation (105) and with constraints (10) or (128).

## 10. Conclusion :

The basic approach in this paper is to place a new method to extend the only available solution of Fox[4] for the H-functions from LNSIE as solution of a Riemann Hilbert problem. The representation of H- functions obtained from LNSIE and from LNIE has become the same. The equivalence between application of the theory of linear singular integral equation and application of Wiener-Hopf technique to the linear integral equations is proved to be true so far it relates to the H-functions .The numerical evaluation of these H-functions from this new form is awaiting for communication. This new method may be applied in anisotropic line transfer problems in non coherent scattering, in problems of multiple scattering and also in time dependent problems of radiative transfer to determine the H – function related to those problems.

**Acknowledgement**:

I express my sincere thanks to the Department of Mathematics , Heritage Institute of Technology , Anandpur , West Bengal , India for the support extended.

-------------------------------------------------